\documentclass{article}

\usepackage{amssymb}
\usepackage{color}
\usepackage{amsmath}

\usepackage{natbib}

\usepackage{graphicx} 

\def\suml#1#2{\sum\limits_{#1}^{#2}}

\begin{document}

\title{Forecasting the underlying potential governing  the time series  of a dynamical 
system}

\author{V.~N.~Livina$^{1,2}$, G.~Lohmann$^3$, M.~Mudelsee$^{3,4}$, and T.~M.~Lenton$^5$}

\date{\small\it $^1$National Physical Laboratory, Teddington, UK\\
$^2$University of East Anglia, Norwich, UK\\
$^3$
Alfred Wegener Institute for Polar and Marine Research, Bremerhaven, Germany\\
$^4$
Climate Risk Analysis, Hannover, Germany\\
$^5$
College of Life and Environmental Sciences, University of Exeter, UK
}

\maketitle

\begin{abstract}
We introduce a technique of time series analysis, potential forecasting, which is based
on dynamical propagation of the probability density of time series. 
We employ polynomial coefficients
of the orthogonal approximation of the empirical probability distribution and extrapolate
them in order to forecast the future probability distribution of data. 
The method is tested on artificial data, used for hindcasting observed climate data, 
and then applied to forecast Arctic sea-ice time series. 
The proposed methodology completes a framework for `potential analysis' of tipping points  
which altogether serves anticipating, detecting and forecasting  
 non-linear changes including bifurcations using several independent techniques of time series analysis.
 Although being applied to climatological series in the present paper, the method
is very general and can be used to forecast dynamics in time series of any origin.
\end{abstract}

\section{Introduction}

Many dynamical systems in general, and geophysical subsystems in particular,
lack analytical deterministic descriptions with fully developed physical models, being
represented mainly by recorded time series. 
At the same time such systems may be of great public interest and societal impact, 
such as the current climate change with rising temperature records around the globe. 


In these circumstances, powerful research tools may be provided by 
statistical time series analysis~\citep{priestley,tong,box}, 
which have found entry into the analysis of climate data 
from a general viewpoint~\citep{manfred} and also from a nonlinear 
dynamical system viewpoint~\citep{donner}. In particular, 
system dynamics can be approximated by means of simple generalised 
stochastic models, where uncertain or unknown variables are represented 
by stochastic components~\citep{hasselmann,chan}. Of 
specific regard to the present paper are time series analysis methods 
that deal with correlations and scaling of 
fluctuations~\citep{peng,podobnik,laloux,plerou,eva,fraedrich}.

In previous papers~\citep{livina07,livina10,ClimDyn}, we have developed
several time series techniques for anticipating and
detecting 'tipping points' in trajectories of dynamical systems, with
applications in climatology. Modified 'degenerate
fingerprinting'~\citep{livina07} was introduced for early warning of
critical behaviour in time series to allow one to 
anticipate an upcoming bifurcation or transition (climate tipping points~\citep{lenton08}).
 This was based on 'degenerate fingerprinting'~\citep{held}, where
the decay rate in the series is monitored using lag-1 autocorrelation in an autoregressive model
(AR1). Modified degenerate fingerprinting employs Detrended Fluctuation Analysis for the
same purposes.
For noisy time series,
we developed  the method of potential analysis~\citep{livina10,ClimDyn},
which  derives the number of system states under
 the assumption of quasi-stationarity of  a data subset. 
This can distinguish a transition, 
which may be a forced drifting of the record without structural changes
in the fluctuations, from a bifurcation which happens when the underlying system potential (the system states
that a climate variable may sample) changes in structure, e.g. instead of two potential wells 
one or three wells appear. A bifurcation is characterised by structural change in the 
dynamical system, whereas transitional series preserve the same structure of fluctuations.

If both techniques give indication of dynamical change, this denotes
a genuine bifurcation. If modified degenerate fingerprinting
indicates a change but potential analysis does not, this  means
a transition rather than a bifurcation, with no changes in
the underlying  system potential.

In this paper, we develop the methodology further, so that we become able
 to not only anticipate and detect, but also  to forecast
the time series dynamics. The skill of such a forecast will depend on several
factors, in particular, whether the upcoming change
will be gradual or abrupt, at what rate it will be happening and how
the scaling properties of the stochastic component may change with time.

Here we outline the methodology, 
test it on artificial data, in several hindcast case studies, and provide
a forecast of the dynamics of Arctic sea-ice extent in the nearest future.

\section{Methodology}

\subsection{Potential analysis as the basis of the method}

We consider a simple stochastic model with a polynomial
potential $U$ as an approximation of the system dynamics,
\begin{equation}
\dot x(t) = -U'(x) +\sigma \eta,
\label{eq_pot}
\end{equation}
where  $\dot x$ is the time derivative of the system variable  $x(t)$ (time series of
an observed variable), $\eta$ is Gaussian white noise of unit variance and $\sigma$
is the noise level.
In the case of a double-well potential, it  can be approximated by a polynomial of 4th order:
$$
U(x)=a_4x^4+a_3x^3+a_2x^2+a_1x.
$$

According to the Fokker-Planck equation for the dynamic evolution of 
the probability density function  $p(x,t)$,
\begin{equation}
\partial_t p(x,t)=\partial_x [U'(x)p(x,t)]
+\frac{1}{2}\sigma^2\partial_x^2 p(x,t)
\end{equation}
its stationary solution is given by~\citep{gardiner}
\begin{equation}
p(x)\sim \exp[-2U(x)/\sigma^2].
\end{equation}
The potential can be reconstructed from
time series data of the system as
\begin{equation}
U(x)=-\frac{\sigma^2}{2}\log p_{\rm d}(x),
\label{eq_potential}
\end{equation}
which means that the empirical probability density $p_{\rm d}$ has the number
of modes corresponding to the number of wells
of the potential.

This simple approximative approach allowed us to reconstruct the
system potential of various climatic records
(see~\citep{ClimDyn}). It works with remarkable accuracy for data subsets
of length as short as 400 to 500 data points, demonstrating
above 90\% rate of accurate detection, as was shown in an experiment with
artificial data. For data subsets of length above 1000 points
it correctly detects the structure of the potential with  a rate of 98\% \citep{ClimDyn}.
 Potential analysis was introduced in~\citep{livina10}
which is an open-access paper published by Copernicus.org; this makes it
easily accessible for the broad readership and more details on the methodology can be found there.

Here we develop the potential method beyond its detection capability,
such that we are able to forecast the behaviour
of a time series on the basis of its potential. To that effect,
we introduce an extrapolation technique that would use the potential structure
of the time series with linear extrapolation of the coefficients of the approximating polynomials.
To reduce the biases introduced during various stages of the potential analysis 
(due to kernel distribution approximation,
further logarithmic transformation, noise estimation, and finally polynomial fits),
we use the empirical probability density rather than its
logarithmic transformation, the potential (see Eq.~\ref{eq_potential}).
Moreover, unlike in previous work~\citep{ClimDyn},
we use not the $2N$ potential coefficients (where $N$ is the number of potential wells)
but the coefficients of the approximation
of the empirical probability density by a finite Chebyshev polynomial series
(following the approach of~\citep{neagoe}).
Chebyshev approximation has an advantage of being near optimal, 
and already 10th-degree approximation in most
cases of observed time series provides an accurate fit with low values of  the
error function.

\subsection{Approximation of the probability density}

To approximate the empirical probability density, we use the orthogonal (in the interval $[-1,1]$)
Chebyshev polynomials of the first kind:
\begin{equation}
\begin{array}{l}
T_0(x)=1,\\[10pt]
T_1(x)=x,\\[10pt]
\dots \\[10pt]
T_{n+1}(x)=2xT_n(x)-T_{n-1}(x).
\end{array}
\end{equation}
\label{eq_chebyshev}
The polynomial $T_n(x)$ has $n$ zeros in the interval $[-1,1]$ at points
$$
x=\cos\left(
\frac{\pi(k-1/2)}{n}\right), \quad k=1,2,\ \dots , \ n.
$$

The approximation of a function $f(x)$ can be done by using a truncated (finite $N$)
sum of the following form
\begin{equation}
\label{eq_approx}
f(x)\cong\left(\suml{k=0}{N-1}c_kT_k(x)\right)-\frac 1 2 c_0,
\end{equation}
where  $c_k$ are the coefficients obtained by discrete cosine transform of the vector of
nonuniformly spaced samples of the considered function over the sampling grid 
$$
x_k=\cos\left(\frac{\pi(k-1/2)}{n}\right), \quad k=1,2,\, \dots\, , n.
$$ 
For an arbitrary interval $[a,b]$ it is necessary to transform variables as 
$$
y=\frac{x-0.5\cdot(b+a)}{0.5\cdot (b-a)}.
$$  
A good example of the above calculations is given in~\citep{neagoe}.

When decomposition~(\ref{eq_approx}) is obtained according 
to the particular time series problem to be analysed,
the resulting polynomial
is expanded, thus producing the final coefficients
\begin{equation}
f(x)\cong\left(\suml{k=0}{N-1}\tilde C_k x^k\right).
\end{equation}
\label{eq_coefs}

\subsection{Linear extrapolation of the coefficients and forecast time series}

We consider the approximation of the empirical probability density using Chebyshev polynomials
$T_0,\,\dots \, ,T_{10}$. When the approximating polynomial is derived, the
decomposition coefficients
$\tilde C_k$  (Eq.~7) are linearly extrapolated using a set of preceding values. The interval
of these, as well as the extension of the extrapolated 
interval can be  chosen according to the particular
time series to be analysed.  If the series dynamics is homogeneous and steady, the
horizon of the forecast may be as far as several decades of daily data (i.e., thousands of 
points). If the dynamics is highly nonlinear and abrupt, the horizon of the forecast 
may be very short, and the forecast probability density function may differ significantly from 
the observed one at the end of the forecast interval. However, even in this case, although the 
pdfs may differ, the simulated time series --- due to the sampling in the close domain 
and sorting according to the observed data ---  is realistically similar to the hindcast data
(see the example of artificial data in Fig.~\ref{ad_forecast}). As a rule of thumb, the horizon
of forecast may be chosen at the scale of several hundreds up to several thousand points, which 
in case of daily data means up to 10 years of forecast horizon.

The initial pdfs (those providing sequences of potential coefficients to be extrapolated) 
are estimated for fixed time interval, in a sequence of subsets prior to the 
forecast starting point (in sliding windows). 
The extrapolation of parameters provides adequate scaling of the pdf, and no further normalisation 
is necessary.

Once a new probability density is calculated, we generate 
a forecast time series using  a rejection sampling
algorithm~(see, for instance,~\citep{gilks}). This provides 
an artificial series with the prescribed distribution,
but this may be not enough for obtaining a realistic forecast 
time series, because the ordering of the series (and
hence scaling properties like long-term memory) should be 
reconstructed according to the initial data. For this
purpose, we apply so-called "sorting" of  the time series, 
 that means arranging its values in the same order as
in the initial data  (before the forecast started), 
thus reproducing realistic correlations (because their distributions are already
very similar due to the extrapolation of probability density).
Sorting is a simple numerical algorithm which uses 
ranking of the values of two series,
initial subsample and forecast subsample. 
However, this should be done with care, especially in data with
seasonality: if there is a seasonal trend, it is 
very important to sort the forecast series according
to observed data at the same date of the year, 
so that further seasonal variability would be adequate. This
is achieved by going back along the series with  a step 
equal to the seasonality period (365 for daily data or 12
for monthly data). Since certain years may be anomalous in
fluctuations  (due to internal variability in the system),
the initial data used for sorting
may be an average over several years starting from the 
same date in a year (for instance, March 1st in
several  consecutive years). This average is then used 
to sort the forecast series starting on March 1st and projecting
into  the future.

\subsection{Uncertainties and applicability; criteria of performance}

It is necessary to note that minor uncertainties are introduced at various steps of the 
forecasting algorithm: first when  the potential stochastic model is used as
an approximation of dynamics, then when the empirical probability density is 
approximated by Chebyshev polynomials (minor outliers); furthermore,
the polynomial coefficients are linearly extrapolated, which means that the actual
dynamics is correctly forecast only in the case of linear evolution of such coefficients.

In many cases of abrupt highly nonlinear dynamics the linear extrapolation of the
decomposition coefficients may produce  an 
empirical distribution with large deviations, especially
in case of non-stationarity of the data. We attempted bootstrapping of the
decomposition of coefficients according to~\citep{manfred}. Based on
bootstrapping techniques, it is possible to consider blocks of data in a chosen subset $x$
of size 
\begin{equation}
L=NINT\left(W^{1/3}\frac{\sqrt{6}a_1(x-\bar x)}{1-(a_1(x-\bar x))^2}\right),
\label{eq_manfred}
\end{equation} 
where  $NINT(\cdot)$ is the nearest integer function, 
$W$ is the window length, $a_1$ is the lag-1 autocorrelation, $\bar x$ is the mean
value of the subset $x$.
This block length selector was derived in~\citep{manfred} from~\citep{sherman},
who adapted a formula from~\citep{carlstein}. For 
$a_1\to 0$, $L$ is chosen equal to 1; when the denominator of Eq.~(\ref{eq_manfred}) 
tends to 0, $L$ is chosen equal to $N-1$.

In the case of nonstationary data, when the probability distribution varies within the
data subset, bootstrapping provides estimates of the partial probability  distributions, 
which may deviate from the average quite significantly. In practical terms, applying
 bootstrapping for estimation of the decomposition coefficients in non-stationary data
provided worse results in the considered samples, 
with the forecast time series of poorer skill than the single-estimated
probability density functions.  A possible solution to this could be modification of the
bootstraping algorithm, where instead of mean value removal a more sophisticated detrending
is applied.  We plan to adapt block bootstrap methods~\citep{manfred} (Chapter~3 therein) for 
that means in a future study.

Furthermore, the important parameter that affects the skill of the forecast is the
extrapolation period. The skill of the forecast drops with increase of its value.

Obviously, in the case of abrupt changes using linear extrapolation of coefficients
may prove unsatisfactory, and our method in those cases may be not applicable. The best
results are obtained when the data undergoes gradual dynamic change and the forecast horizon is
within 100 time units (which means 3 months for daily data and up to 8 years for monthly data).

To assess the skill
of the forecast, we used several techniques widely  
applied in modelling community for comparison of
observed and modelled data, for instance,
in  hydrology~\citep{nash,joh}:
$$
\begin{array}{ll}
\mbox{Daily Root Mean Square} \ \ \ \ &
 \sqrt{\frac 1n\sum_{i=1}^N(x_m^i-x_o^i)^2}, \\[10pt]
\mbox{Nash-Sutcliffe efficiency} \ \ \ \ &
1-\left(\sum_{i=1}^N(x_m^i-x_o^i)^2\left/
\sum_{i=1}^N(x_o^i-\overline{x})^2\right.\right),\\[10pt]
\mbox{Percent bias} \ \ \ \ &
\left[\sum_{i=1}^N(x_m^i-x_o^i)\left/\sum_{i=1}^N(x_o^i)\right.\right]
\times 100,
\end{array}
$$
where $x_o$ and $x_m$ are the observed and
modelled  series, respectively, $i=1,\,\dots,\,N$ is
time index, $\overline{x}$ is the
mean value of the series $x$.
It is easy to see that for two
identical time series DRMS=0, NS=1, and \%bias=0, and any deviation from those values would
indicate the difference between the modelled and observed time series pointwise.
 
However, it is necessary to mention that these skill estimators perform best when applied to 
data with normal distribution without outliers, 
which is not the case in most of our datasets; therefore, they may 
introduce additional biases in estimation of accuracy of 
arbitrary data; we apply them for general information
only.
                                  
\section{Tests of artificial and climate data}

\subsection{Artificial data}

\begin{center}
\begin{figure*}[h!]
\includegraphics[width=\linewidth]{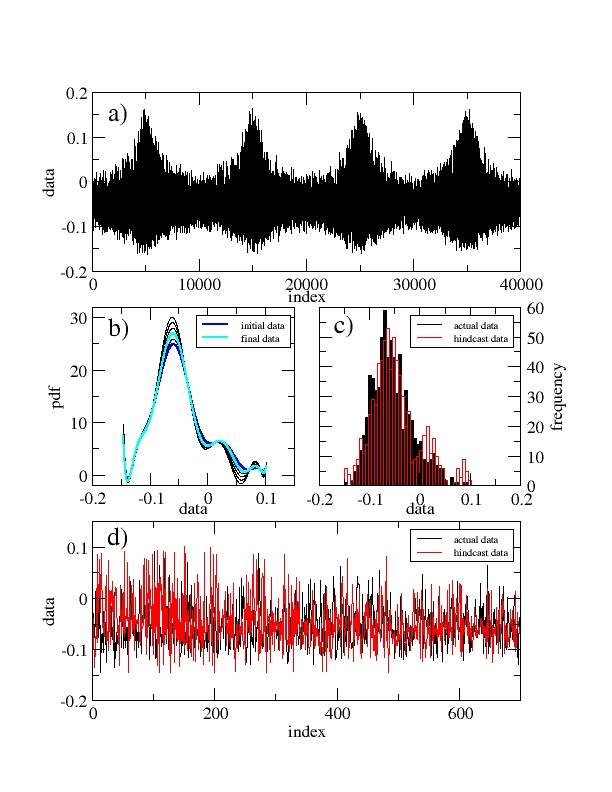}
\caption{Artificial data with oscillating potential, from one to 
two wells and back: a)~time series; b)~hindcast empirical
probability density 
(Chebyshev-polynomial approximations, which may result in some negative pdf values),
where blue curve is the initial statistics 
at the beginning of the hindcast, black curves are extrapolated
densities up to 100 time units ahead, cyan curve is the real pdf 
at the end of the forecast, for comparison with extrapolation;
c)~histograms of the forecast and real data at the end of extrapolation; 
d)~time series corresponding to histograms in the panel c).}
\label{fig_andrea}
\end{figure*}
\end{center}

\begin{center}
\begin{figure*}[h!]
\includegraphics[width=\linewidth]{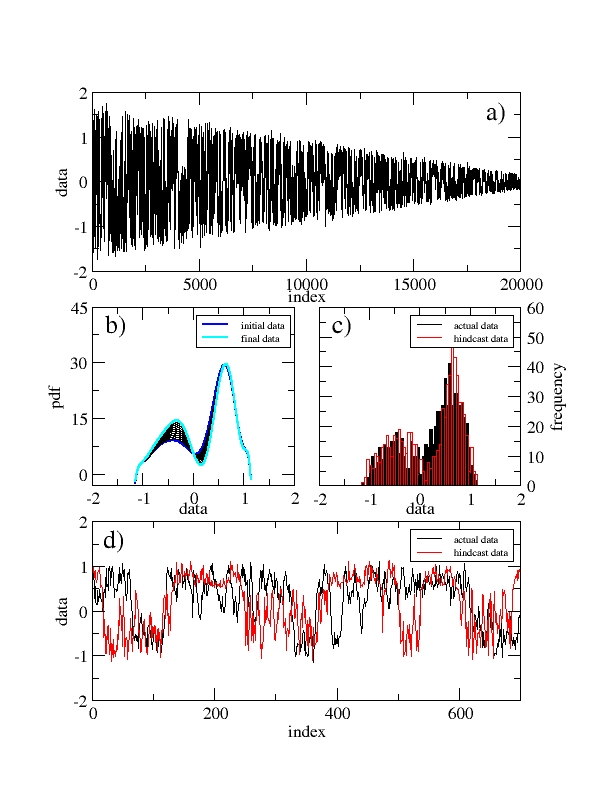}
\caption{As in Fig.~1, for artificial double-well-potential data with decreasing noise level.}
\label{fig_decr}
\end{figure*}
\end{center}

\begin{center}
\begin{figure*}[h!]
\includegraphics[width=0.7\linewidth]{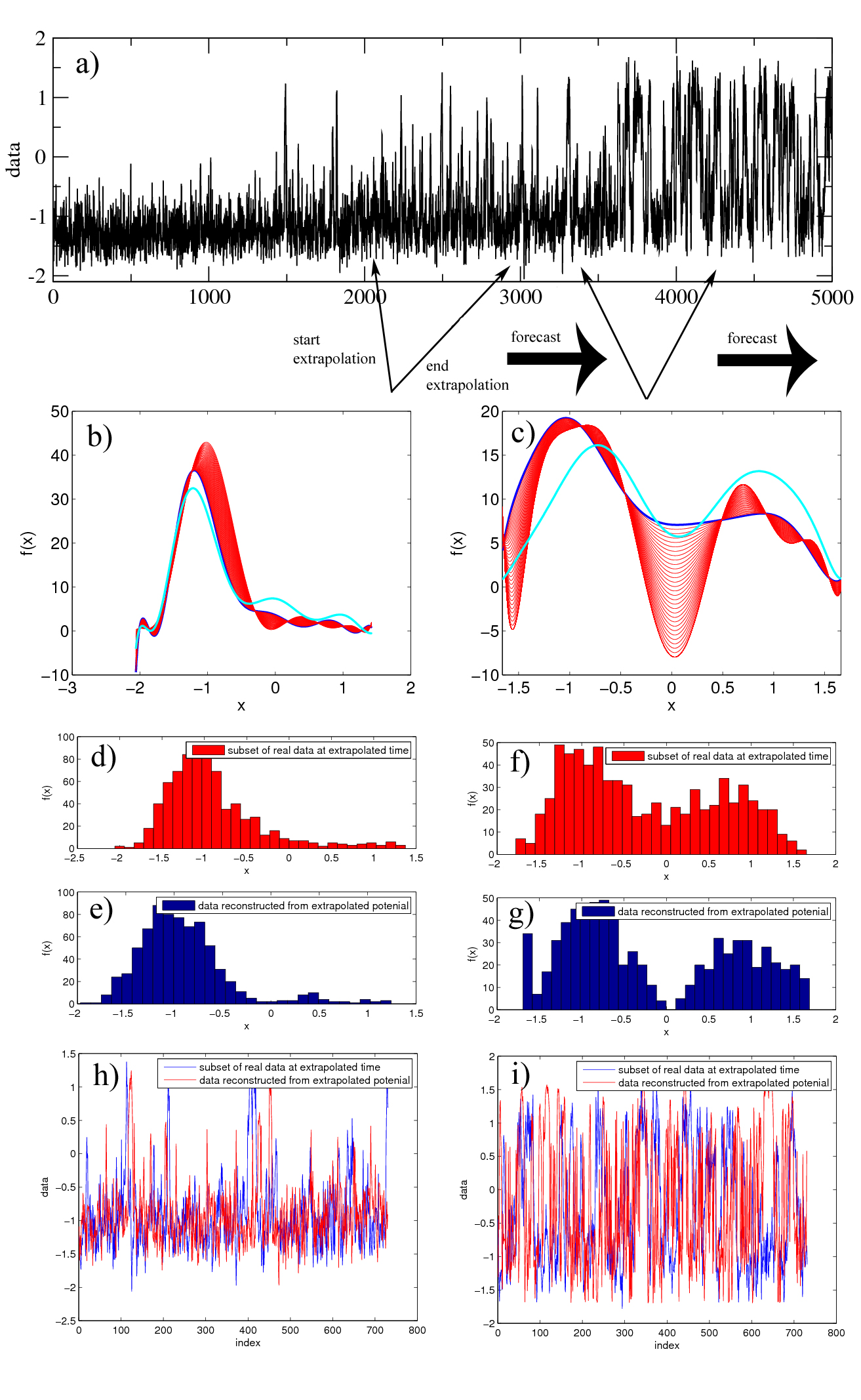}
\caption{Artificial data bifurcating from one-well to double-well dynamics --- two hindcasts 
are demonstrated.
The first hindcast, from point 2100, has initial potential ending at 2100 and the
final potential (for the hindcast comparison) 
starting after 800 points of extrapolation, that is, at 2900. Start and end of the extrapolation
are indicated by two thin arrows under the top panel. 
After this, the forecast is running for 730 days
ending at 3630, which is indicated by the thick line underneath. Panels:
a)~time series -- the dynamics is potentially 
extrapolated in two intervals: from 2100 to 2900
and from 3400 to 4200; 
b)~emprical probability density (Chebyshev-polynomial approximations) for the
interval from 2100 to 2900: blue curve is the initial probability density, 
cyan is the final curve; red curves are
extrapolated at equal steps; c)~the same as~(b)~for the interval from 3400 to 4200; 
(d) and~(e)~are histograms for extrapolation
at point 2900: comparison of the actual histogram 
and the result of extrapolation; (f) and~(g)~are the same as~(d) and~(e)~for
the point 4200; h)~comparison of the forecast and 
actual data in interval from 2900 to 3630: i)~comparison of the
forecast and actual data in interval from 4200 to 4930.
}
\label{ad_forecast}
\end{figure*}
\end{center}

We considered several simulated time series:
an artificial dataset where the potential is varying from single-well
to double-well and back several times (Fig.~\ref{fig_andrea}),
double-well-potential data with decreasing noise level (Fig.~\ref{fig_decr})
and artificial data bifurcating from one-well- to double-well-potential 
(simulated tipping point, Fig.~\ref{ad_forecast}).  Our aim is to test if the 
proposed methodology is capable to capture the modelled dynamics of the series 
and forecast the record adequately.

We perform hindcast of these series by choosing a certain point where we
start extrapolation of the empirical probability density,
then we compare  the modelled series with actual data at the end of the
forecast.  Figures~\ref{fig_andrea} and~\ref{fig_decr} show datasets in panels (a), 
the probability density functions and histograms of the data and hindcast 
in panels (b,~c) of each plot,
and the samples of  modelled data in panels~(d).

Figure~\ref{ad_forecast} combines two hindcasts of the series bifurcating from one-well
to double-well potential, in the intervals shown by arrows in the figure 
(probability density functions and histogram of observed and modelled data in 
panels (b,~d,~e,~h) and (c,~f,~g,~i)). 
The initial
potential is  $U(x)=x^4-2x^2$; then the term with 1st power of $x$ starts
growing gradually from zero
until the potential reaches form  $U(x)=x^4-2x^2+8x$. 
In this experiment, the forecast is performed in total for 
1530 ``days'' without any intermediate assimilation
of modelled data, and for these  conditions the modelled 
series is very close in statistical properties 
to the hindcast data (although the probability density 
is not entirely identical). The result demonstrates
that for certain bifurcating systems with gradual 
dynamics the methodology may be very efficient as a 
forecast tool. 

\subsection{Climate data}

\begin{center}
\begin{figure*}[h!]
\includegraphics[width=\linewidth]{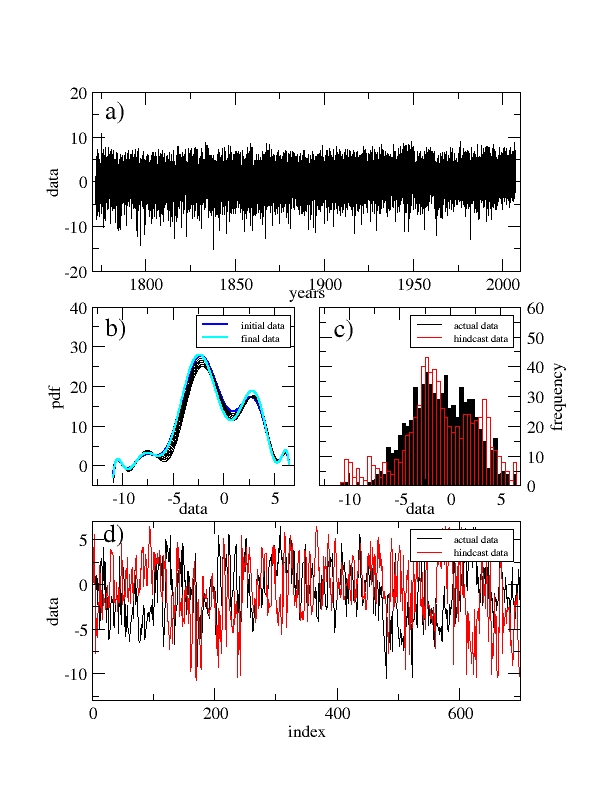}
\caption{Central England Temperature deseasonalised 
fluctuations: a)~time series; b)~hindcast empirical
probability density (Chebyshev-polynomial approximations), where blue curve is the initial 
statistics at the beginning of the hindcast, black curves are extrapolated
densities up to 100 time units ahead, cyan curve is 
the real pdf at the end of the forecast, for comparison with extrapolation; 
c)~histograms of 2-year-long hindcast: the forecast and real data at the end of extrapolation; 
d)~time series corresponding to histograms in the panel (c).}
\label{fig_cet}
\end{figure*}
\end{center}

\begin{center}
\begin{figure*}[h!]
\includegraphics[width=\linewidth]{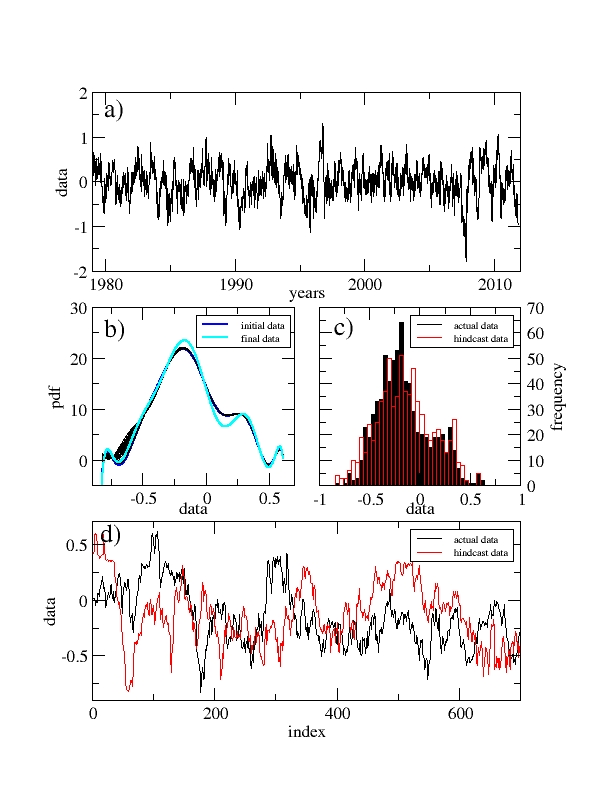}
\caption{As in Fig.~4, for Arctic sea-ice area fluctuations after 
deseasonalising and removal of quadratic decreasing trend.}
\label{fig_uoi3600}
\end{figure*}
\end{center}

Similarly to artificial data in the previous section, 
we performed hindcast experiments with observed
climate data, temperature and sea-ice extent. The results 
are shown in Figures~\ref{fig_cet},\ref{fig_uoi3600}.

The Central England Temperature~\citep{manley,parker},  which is available as monthly series
since 1659 and as daily since 1772, is considered here in daily format. It is
first deseasonalised  (subtracting the average annual cycle of daily data) and then
the hindcast is performed as in the above artificial records.

Sea ice area data were obtained from `The Cryosphere Today' project of the University of Illinois.
This dataset\footnote{http://arctic.atmos.uiuc.edu/cryosphere/timeseries.anom.1979-2008} 
uses SSM/I and
SMMR series satellite products and spans 1979 to present at daily resolution. 
The fluctuations of the Arctic sea-ice area are not only deseasonalised,
but also the quadratic trend is removed, as we pre-processed the
sea-ice data in our recent paper~\citep{cryo}.  This was done to study the
properties of the fluctuations.

Altough the real data has much more complicated variability and dynamics, the method performs
as well as in the cases of artificial data analysed above, with modelled series having the same 
statistical properties as the initial data.

\subsection{Forecast skill}

\begin{center}
\begin{figure*}[h!]
\includegraphics[width=\linewidth]{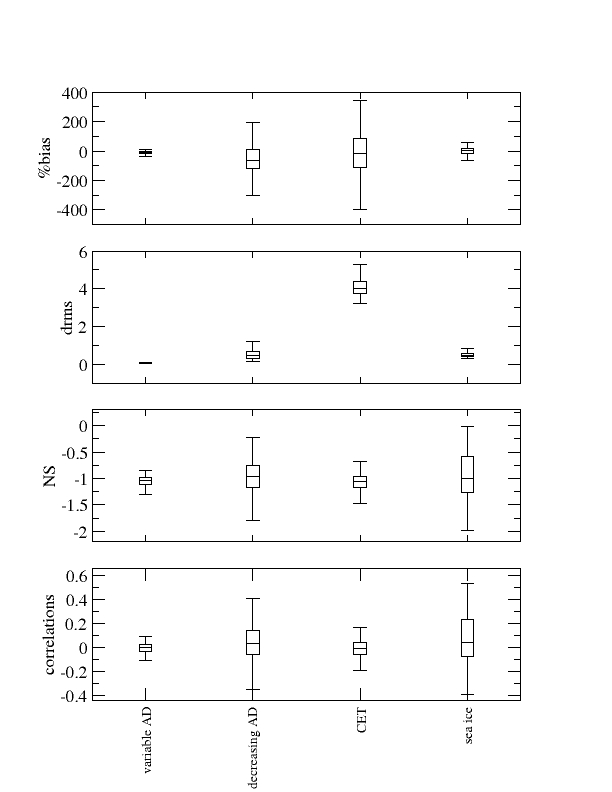}
\caption{The skill statistics of the forecasts of four datasets. Outliers are not shown.
'Variable AD' is the dataset analysed in Fig.~1; 'Decreasing AD' is the dataset analysed in 
Fig.~2; 'CET' is the dataset analysed in Fig.~4; 'sea ice' is the dataset analysed in Fig.~5.}
\label{fig_skill}
\end{figure*}
\end{center}

The skill of the forecast is calculated for multiple subsets along
series using the hindcast approach as described above. We show 
the skill in Figure~\ref{fig_skill}  as boxplots with whiskers (outliers are not 
shown) and conclude that in some cases
the exact statistical properties and correlations are not reproduced well, 
with acceptable mean values over
subsets but rather large standard deviations.
This is because our stochastic approach is based
on the probability density function, which means that the exact
dynamics may vary for series with the same histograms, and skill estimators based on point-wise
comparison of time series may deviate from ideal values.
However, the patterns of the series are very close and suitable for long-term
stochastic predictions, as we show below for the Arctic sea-ice.

The stochastic forecast is not expected to be precise at particular 
values. For example, even the sophisticated deterministic climate models still struggle
to forecast actual weather at medium-term scale. Our forecast 
technique is much lighter computationally, being based on a simple stochastic model; yet 
it reproduces the pattern of the series and takes into account typical seasonal variability
from the past, which produces realistic forecast series. There exist regional climate models
that achieve good accuracy of forecast, but those usually implement assimilation of observational
data, which cannot be done when attempting longer-timescale forecast.

\section{Future Arctic sea-ice area dynamics}

Arctic sea-ice dynamics have been a topic of recent scientific debate, 
and available estimates of when summer ice cover will disappear 
range from as early as 2016 to never;
 the early ice loss estimate comes from extrapolation of PIOMAS data~\citep{zhang,wang,boe}.

There is also an ongoing  debate about whether tipping points  could occur in the
Arctic region (see~\citep{serreze,tietsche,wassmann,lenton_arctic} and references therein).
 Several studies argue that sea-ice loss is highly reversible and therefore not a 
tipping point~\citep{tietsche}. 
Others suggest that reversible tipping points can occur and summer sea-ice loss may be 
one candidate~\citep{lenton08}. 
Here we restrict ourselves to the question of when summer ice cover is forecast to disappear in the Arctic.

Here we propose 
a stochastic forecast based on the above methodology, 
without taking into account complex feedbacks (that may
reverse the declining dynamics as well as speed it up). 
Assuming that the present dynamics  continues its gradual
development, and taking into account what we already 
know about this time series, we forecast it
using potential analysis.

\begin{center}
\begin{figure*}[h!]
\includegraphics[width=\linewidth]{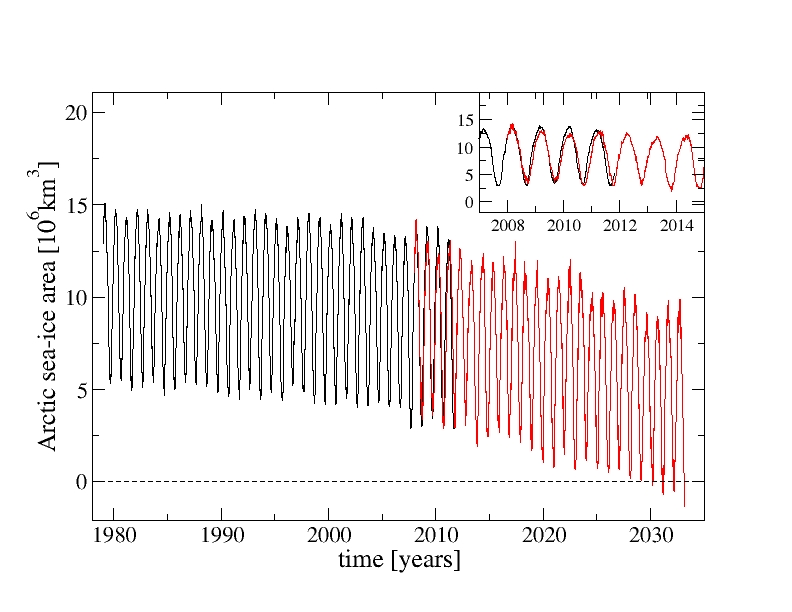}
\caption{Arctic sea-ice area forecast until 2035, which indicates zero 
level of summer sea ice in the 2030s. The inset plot shows magnification
of the main plot for the period 2007-2015, where one can
see how the forecast series (red)
is started from 2008 and extends beyond the observed data (black).}                                               
\label{ice_forecast}
\end{figure*}
\end{center}

In Fig.~\ref{ice_forecast}, we show a long-term forecast of the Arctic
sea-ice by combining our knowledge about seasonality and
current trends in the data. From the point of forecast (which is chosen
in 2008 rather than most recently in order to assess the
accuracy of the forecast over the recently observed data),
we combine the last 5-year seasonal average, extrapolated quadratic trend,
and the forecast of the fluctuations as described above.
In the inset we show that between 2008 and 2010 the forecast is surprisingly
good and very close to the observed data; the later departure
is mainly due to very anomalous summers in the recent few years.

The forecast indicates complete loss of Arctic  summer sea-ice area by  the 2030s.
Obviously, the method provides an estimate of sea ice loss in the Arctic Ocean
if the system will not experience new feedbacks which are not included in our approach.
It is therefore a conservative estimate solely based on extrapolation of the current trends.

\section{Discussion.}

We have developed a forecasting technique based on potential analysis
with dynamical propagation of the probability density. 
As the main idea, we employ approximations of the empirical 
probability  distributions and extrapolate them to the 
future probability distribution of data.
We conducted
experiments with artificial, model and observed climatic data and showed the
efficient forecast performance of the new methodology.

As one particular example, we applied the method to the 
expected sea ice trend in the Arctic Ocean.
The dynamics of Arctic sea ice was forecast and shown to  disappear in summer in the 2030s if
the current trends remain the same. This  can be compared to fully dynamical models.

As a logical next step, bootstrap resampling (Section 2.4)  can help to construct prediction
 confidence intervals based on percentiles or standard errors~\citep{manfred}. 

\bigskip

\noindent
{\bf Acknowledgement.}
The idea of the paper was developed during VNL's visit to Alfred
Wegener Institute for Polar Research, Germany (February 2012).
VNL  and TML were supported by NERC through the
project "Detecting and classifying bifurcations in the climate
system" (NE/F005474/1) and by AXA Research Fund through a
postdoctoral fellowship.
MM is supported by the European Commission as PI in the Marie Curie
Initial Training Network LINC (project number 289447) under the Seventh
Framework Programme.
The Central England Temperature dataset was provided by
the British Atmospheric Data Centre (BADC),
which is part of the NERC  National Centre for Atmospheric Science (NCAS). 
The research was carried out on the High
Performance Computing Cluster supported by the Research Computing Service at
the University of East Anglia.


\end{document}